\documentclass[a4paper,11pt]{article}
\usepackage{amsmath, amssymb, amsfonts,indentfirst,graphicx,color,anysize}
\marginsize{3cm}{3cm}{2cm}{2cm}
\makeatletter
\DeclareRobustCommand{\cev}[1]{%
  {\mathpalette\do@cev{#1}}%
}
\newcommand{\do@cev}[2]{%
  \vbox{\offinterlineskip
    \sbox\z@{$\m@th#1 x$}%
    \ialign{##\cr
      \hidewidth\reflectbox{$\m@th#1\vec{}\mkern4mu$}\hidewidth\cr
      \noalign{\kern-\ht\z@}
      $\m@th#1#2$\cr
    }%
  }%
}
\makeatother

\title{
	\includegraphics[width=0.35\textwidth]{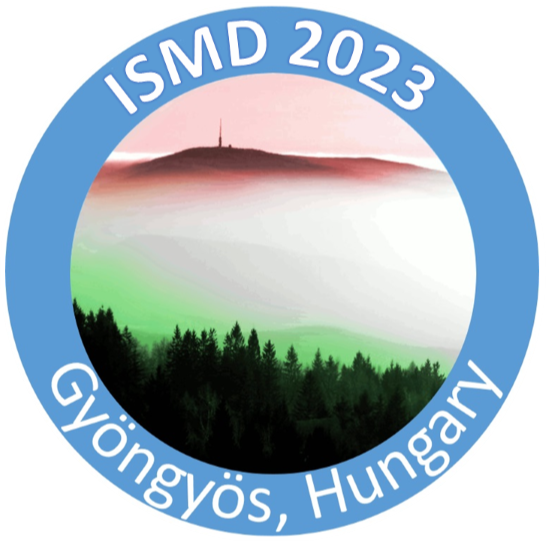}\\[1cm]
	\textbf{Recent Developments of Small-$x$ Evolution for Quark and Gluon Helicity}}
\author{{Yossathorn Tawabutr$^{1,2}$}\\[1ex]
	$^1$Department of Physics, University of Jyv\"askyl\"a, P.O. Box 35, 40014\\University of Jyv\"askyl\"a, Finland\\
	$^2$Helsinki Institute of Physics, P.O. Box 64, 00014 University of Helsinki, Finland\\
}
\begin{document}

\maketitle

\begin{abstract} 

Helicity of quarks and gluons inside the proton at small Bjorken $x$ is one of the missing pieces of the proton spin puzzle, with limited experimental results due to various challenges. To address the problem, we derive under the framework of color glass condensate (CGC) effective theory a renormalization group equation in rapidity for the parton helicity distributions, which resums double-logarithmic factor, $\alpha_s\ln^2(1/x)$, with $\alpha_s$ being the strong coupling constant. With running coupling, the equation produces results at Bjorken $x\leq 0.1$ that are consistent with the world polarized scattering data. There is an evidence for a significant parton helicity contribution from the small-$x$ region. However, the helicity estimate still contains a large uncertainty that will dramatically reduce with the upcoming measurements from the Electron-Ion Collider (EIC). Ongoing attempts to reduce the uncertainty of our helicity prediction in the medium term will also be discussed.

\end{abstract}

\section{Introduction}

This article is directly based on \cite{Cougoulic:2022gbk,Adamiak:2023okq,Adamiak:2023yhz}. See also \cite{Kovchegov:2015pbl,Kovchegov:2016weo,Kovchegov:2016zex,Kovchegov:2017jxc,Kovchegov:2017lsr,Kovchegov:2018znm,Kovchegov:2020hgb,Adamiak:2021ppq} for the preliminary calculations and results leading up to those works.

Proton spin puzzle is a longstanding problem in theoretical physics, concerning the contributions to the spin of a proton coming from the quarks and gluons inside. The problem arose in late 1980s as the European Muon Collaboration (EMC) measured the longitudinal spin asymmetry in the muon-proton polarized deep-inelastic scattering (DIS) process and discovered that the total quark spin inside the proton did not add to the total proton spin of $\frac{1}{2}$ \cite{EMC1, EMC2}. This discovery implies nontrivial amounts of contributions to the proton spin from (i) parton orbital angular momenta (OAMs), (ii) spin of gluons and (iii) spin of quarks that require scattering processes with energies beyond that of the EMC in order to observe. This article focuses on contributions (ii) and (iii) in the helicity basis. 

To systematically study the proton spin, we adopt the Jaffe-Manohar sum rule \cite{JM}, 
\begin{align}\label{JM}
&\frac{1}{2} = S_q + S_g + L_q + L_g\,,
\end{align}
where $S_q$($S_g$) and $L_q$($L_g$) are respectively the spin and orbital angular momenta of quarks(gluons) inside the proton. In the notation of Eq.~\eqref{JM}, the focus of this article is in the first two terms on the right-hand side. These terms can be written further as
\begin{subequations}\label{hPDF}
\begin{align}
S_q(Q^2) &= \frac{1}{2} \int\limits_0^1dx \, \Delta\Sigma(x,Q^2)\,,\label{hPDF_q}\\
S_g(Q^2) &=  \int\limits_0^1dx \, \Delta G(x,Q^2)\,,\label{hPDF_g}
\end{align}
\end{subequations}
where $x$ is the Bjorken $x$ variable and $Q^2$ is the virtuality. Here, $\Delta G$ is the gluon helicity-dependent parton distribution function (hPDF), which is defined to be the difference between the parton distribution function (PDF) of gluons with positive helicity (aligned with that of the target) and that of negative-helicity gluons,
\begin{align}\label{hPDF_g2}
&\Delta G(x,Q^2) = G(x,Q^2,h=+1) - G(x,Q^2,h=-1)\,.
\end{align}
Furthermore, in Eq.~\eqref{hPDF_q}, the function, $\Delta\Sigma$, is the ``flavor singlet'' quark hPDF, which is defined as
\begin{align}\label{hPDF_fsinglet}
&\Delta\Sigma(x,Q^2) = \sum_{q=u,d,s}\left[\Delta q(x,Q^2) + \Delta\bar{q}(x,Q^2) \right],
\end{align}
where throughout this article we only include the three lightest quark flavors. Then, $\Delta q$ and $\Delta\bar{q}$, which are respectively the quark and antiquark hPDFs, are defined in the similar fashion as the gluon counterpart in Eq.~\eqref{hPDF_g2}. In turn, we also define the ``flavor non-singlet'' quark hPDF as
\begin{align}\label{hPDF_fnonsinglet}
&\Delta q^-(x,Q^2) =\Delta q(x,Q^2) - \Delta\bar{q}(x,Q^2) \,,
\end{align}
for each flavor $q=u,d,s$.

More recent experimental measurements from the Relativistic Heavy Ion Collider (RHIC) found the quark and gluon spin contributions at $Q^2=10$ GeV$^2$ to be \cite{RHIC_spin1, RHIC_spin2}
\begin{subequations}\label{RHIC}
\begin{align}
S_q(Q^2=10\text{ GeV}^2) &\simeq \frac{1}{2} \int\limits_{0.001}^1dx \, \Delta\Sigma(x,Q^2=10\text{ GeV}^2) \in [0.15,0.20]\,,\label{RHIC_q}\\
S_g(Q^2=10\text{ GeV}^2) &\simeq  \int\limits_{0.05}^1dx \, \Delta G(x,Q^2=10\text{ GeV}^2) \in [0.13,0.26]\,.\label{RHIC_g}
\end{align}
\end{subequations}
From Eqs.~\eqref{RHIC}, we see that there are nonzero contributions coming from OAM and/or the region of small Bjorken $x$, as even the upper bounds of the confidence intervals add to a number short of the proton spin of $\frac{1}{2}$. Besides, the lower limits of both integrals are considerably greater than theoretical value of zero, c.f. Eq.~\eqref{hPDF}. This reflects the fact that experimental measurements can only be performed at finite values of Bjorken $x$, with lower $x$'s requiring larger center-of-mass energies. On this end, however, the limit will be significantly improved down to $x_{\min}\sim 10^{-4}$ at the upcoming EIC \cite{EIC,AbdulKhalek:2021gbh,Abir:2023fpo}. 

This inspires the main focus of our research program, which aims to develop a renormalization group equation that allows one to relate quark and gluon hPDFs at small Bjorken $x$ to their counterparts at moderate values of $x$, whose results can be determined experimentally. Ultimately, the evolution equation will complement experimental results in determining parton spins at small $x$. This article summarizes the most recent development of the program, together with the outlook of future calculations and analyses. Section \ref{sect:setup} outlines the setup and derivation \cite{Cougoulic:2022gbk,Kovchegov:2015pbl,Kovchegov:2018znm} of the evolution equation, which results in the small-$x$ asymptotic relations \eqref{asymp_singlet} and \eqref{asymp_nonsinglet} for parton hPDFs \cite{Adamiak:2023okq,Kovchegov:2016zex,Kovchegov:2020hgb}. Then, Section \ref{sect:globalana} highlights important results from our recent global analysis \cite{Adamiak:2023yhz} with polarized DIS and semi-inclusive DIS (SIDIS) data. Finally, we conclude in Section \ref{sect:conc} and discuss potential future projects.

\section{Evolution Equation}\label{sect:setup}

The study of parton helicity at small $x$ begins with the definition of quark \cite{Mulders:1995dh} and dipole gluon \cite{Bomhof:2006dp} transverse-momentum-dependent (TMD) PDFs, $g^q_{1L}(x,k^2_{\perp})$ and $g^{g,\,\mathrm{dip}}_{1L}(x,k^2_{\perp})$, respectively. Integrating each TMD over the transverse momentum, $k_{\perp}$, yields the respective parton hPDF. Taking $x$ to be a small parameter, we expand the expression for each hPDF as a power series of $x$ and keep the largest non-vanishing terms. Similar steps can also be applied to the $g_1$ structure function starting from its definition in term of the polarized DIS cross section. At the end, both hPDFs and the $g_1$ structure function can be written in terms of ``helicity-dependent CGC averaging'' over target states \cite{Cougoulic:2022gbk,Kovchegov:2015pbl} of color traces involving ``polarized Wilson lines'', which are constructed from semi-infinite light-cone Wilson lines \cite{Cougoulic:2022gbk,Kovchegov:2015pbl,Kovchegov:2018znm}. A light-cone Wilson line at transverse position $\underline{x}$ from light-cone time $a^-$ to $b^-$ can be written as \cite{Yuribook}
\begin{align}\label{Vunpol}
V_{\underline{x}} &= \mathcal{P}\exp\left[ig\int\limits_{a^-}^{b^-}dx^-A^+(0^+,x^-,\underline{x})\right],
\end{align}
where $g$ is the strong coupling and $A^{\mu} = \sum_aA^{a\mu}t^a$ is the gluon field of the target with $t^a$ being the generator of the $SU(N_c)$ gauge group and $N_c$ being the number of colors. Here and throughout this article, we use the convention such that a four vector in the light-cone coordinates can be written as $v = (v^+,v^-,\underline{v})$ where $v^{\pm} = \frac{v^0\pm v^3}{\sqrt{2}}$.

Under the framework of the CGC effective theory \cite{Gelis:2010nm}, a fundamental(adjoint) Wilson line corresponds to a projectile quark(gluon) moving in the light-cone minus direction and interacting with the target, which is moving in the light-cone plus direction. The limit of small Bjorken $x$ corresponds to a high center-of-mass energy, that is, the projectile has a large light-cone minus momentum. In this regime, the interaction time scale is much shorter than the lifetime of the minus-moving quark or gluon as a part of the projectile. This leads to the convention of denoting the interaction region by the ``shockwave'' \cite{Balitsky:2001gj}.  

With regard to the large minus momentum of the projectile, the most significant contribution -- the ``eikonal'' contribution -- to the projectile-target interaction consists of multiple exchanges of target gluon fields, $A^+$, as suggested by the expression \eqref{Vunpol} for the Wilson line \cite{Yuribook,Gelis:2010nm,Balitsky:2001gj}. In fact, this is a building block for the unpolarized small-$x$ evolution involving the ``dipole amplitude'', which is the convenient degree of freedom to study DIS and other unpolarized processes under the CGC framework \cite{Gelis:2010nm}. However, this interaction does not know about the helicity of the projectile. 

As hPDFs are probed through longitudinal spin asymmetry, which requires polarized DIS processes \cite{Lampe:1998eu}, we require a structure in the projectile that is capable of delivering the information about the projectile helicity to the target. This requires the inclusion to the Wilson line described above of ``sub-eikonal'' corrections, which are relatively suppressed by a power of projectile's light-cone minus momentum \cite{Kovchegov:2015pbl,Kovchegov:2021iyc,Altinoluk:2014oxa,Chirilli:2018kkw,Chirilli:2021lif}. Diagrammatically, three different categories of interaction are relevant to the study of helicity \cite{Cougoulic:2022gbk,Kovchegov:2018znm}. Each of the three categories involves the usual multiple gluon exchanges at the eikonal level, together with the additional sub-eikonal interaction, which can be the exchange of
\begin{enumerate}
\item[(i)] two quarks,
\item[(ii)] transverse gluon field, $\underline{A}$, with the structure of strong magnetic moment, $F^{12}$,
\item[(iii)] transverse gluon field, $\underline{A}$, in the form of covariant derivatives, $\cev{\underline{D}}\cdot\vec{\underline{D}}$, acting on the eikonal Wilson lines.
\end{enumerate}
Only the first two sub-eikonal exchanges depend explicitly on helicity, while the third category is only important because of the way it convolutes with (i) and (ii) in the resulting helicity evolution equation. Remarkably, with the quark exchange being a significant contribution, one should take the quark degrees of freedom into account when study helicity at small Bjorken $x$, despite the fact that gluons are typically greater in number in this regime \cite{Cougoulic:2022gbk,Kovchegov:2015pbl}. This is in contrast to the unpolarized counterpart \cite{Balitsky:1995ub,Balitsky:1998ya,Kovchegov:1999yj} for which the pure-glue limit suffices. It is convenient to employ the ``polarized dipole amplitudes'', which are polarized CGC averaging of the trace of a polarized and an unpolarized Wilson line, as the degree of freedom for our helicity evolution equation \cite{Kovchegov:2015pbl}.

With the sub-eikonal operators and their corresponding diagrams set up, the small-$x$ evolution equation for helicity can be constructed by re-scaling the shockwave in such the way that an extra emission and absorption of a parton, which were formerly parts of the shockwave, become excluded from the shockwave. Instead, these parton exchanges are taken into account through perturbative QCD calculation. The complete explanation and explicit derivation of the evolution equation is given in \cite{Cougoulic:2022gbk,Kovchegov:2018znm}. Ultimately, this process leads to the Kovchegov-Pitonyak-Sievert--Cougoulic-Tarasov-Tawabutr (KPS-CTT) equation, which is the renormalization group equation in rapidity that allows us to obtain a description of hPDFs and the $g_1$ structure function at small $x$ with a more controlled accuracy \cite{Cougoulic:2022gbk,Adamiak:2023okq,Adamiak:2023yhz}. It should be emphasized that the KPS-CTT equation is a high-energy/small-$x$ evolution equation, in contrast to the polarized DGLAP equation \cite{Gribov:1972ri,Altarelli:1977zs,Dokshitzer:1977sg} that evolves hPDFs with the transverse scale, $\mu^2$.

The KPS-CTT equation is not closed in general. With the extra parton line after each step of evolution, the resulting operator involves one extra Wilson line operator, and hence it is a different operator from the one we started with. This is similar to the Balitsky hierarchy for unpolarized small-$x$ evolution \cite{Balitsky:1995ub,Balitsky:1998ya}. A possible workaround is to take the Veneziano large-$N_c\& N_f$ limit \cite{Veneziano:1976wm}, in which $N_c\sim N_f\gg 1$ and $\alpha_sN_c\ll 1$. As a result, each gluon line can be written as a color-octet quark-antiquark pair with the help of Fierz identity. At the same time, vertices proportional to the number of flavors, $N_f$, e.g. the $g\to q\bar{q}$ vertex, remain innegligible. This feature is essential for helicity evolution in which the quark exchange term remains significant \cite{Cougoulic:2022gbk,Kovchegov:2015pbl}. In the large-$N_c\& N_f$ limit, the KPS-CTT equation is closed and involves convolutions between one of the three types of sub-eikonal polarized dipole amplitudes and the unpolarized dipole amplitude. Overall, the equation is non-linear.

To further simplify the equation, we take advantage of the fact that the KPS-CTT equation resums $\alpha_s\ln^2(1/x)$ \cite{Cougoulic:2022gbk,Kovchegov:2015pbl}, as oppose to the unpolarized BK equation that resums $\alpha_s\ln(1/x)$ \cite{Balitsky:1995ub,Balitsky:1998ya,Kovchegov:1999yj}. Thus, the resummation in the former is more significant than that of the latter for each value of $x$, justifying the approximation that the unpolarized dipole amplitude remains at its moderate-$x$ initial condition. As a result, the large-$N_c\& N_f$ KPS-CTT equation linearizes \cite{Cougoulic:2022gbk,Kovchegov:2015pbl}, allowing for an efficient iterative computation of the small-$x$ asymptotic solution \cite{Adamiak:2023okq,Kovchegov:2016weo,Kovchegov:2020hgb}. In turn, this yields the following small-$x$ asymptotic behaviors for hPDFs and the $g_1$ structure function,  
\begin{align}\label{asymp_singlet}
\Delta\Sigma(x,Q^2) &\sim \Delta G(x,Q^2) \sim g_1(x,Q^2) \sim \left(\frac{1}{x}\right)^{3.43\sqrt{\alpha_sN_c/2\pi}} , 
\end{align}
in the case of $N_f = 3$.

Through a similar construction, the small-$x$ evolution equation can be constructed for the flavor non-singlet quark hPDF \cite{Kovchegov:2016zex}. The only difference in this case is the fact that only the terms carrying flavor information from the projectile to the target contribute. This removes diagrams and simplifies the derivation, resulting in the evolution equation that can be solved analytically, implying the small-$x$ asymptotic of \cite{Kovchegov:2016zex}
\begin{align}\label{asymp_nonsinglet}
\Delta q^-(x,Q^2) &\sim \left(\frac{1}{x}\right)^{\sqrt{\alpha_sN_c/\pi}} ,
\end{align}
for the quark flavor non-singlet hPDF.

\section{Global Analysis}\label{sect:globalana}

In order to determine the predicted spin contribution from small-$x$ quarks and gluons inside the proton, one has to integrate Eq.~\eqref{asymp_singlet} over $x$ from the starting point of KPS-CTT evolution down to zero. However, with any fixed coupling $\alpha_s\gtrsim 0.18$, the $x$-integral of Eq.~\eqref{asymp_singlet} becomes divergent, implying an infinite helicity coming from both quark and gluon at small $x$ \cite{Adamiak:2023okq}. This is unlikely a physical prediction.

Realistically, for sufficiently small $x$, it is no longer valid to neglect the single-logarithmic small-$x$ resummations for either the unpolarized dipole amplitude (through BK equation) or the polarized dipole amplitude (as corrections to KPS-CTT equation). While the BK equation is known at the single-logarithmic order (SLA) \cite{Balitsky:1995ub,Balitsky:1998ya,Kovchegov:1999yj}, only parts of the SLA corrections to the KPS-CTT equation have been derived \cite{Kovchegov:2021lvz}, with the complete calculation still in progress \cite{SLAops}. However, the results from \cite{Kovchegov:2021lvz} already show that the helicity evolution equation at SLA involves convolution integrals of unpolarized and polarized dipole amplitudes. As a result, once all the additional resummations are included, saturation effects in BK evolution are expected to suppress both unpolarized and polarized dipole amplitudes and hence limit the total spin contribution to a finite value.

Despite the limited progress in the SLA corrections, it is possible to cross check the KPS-CTT equation with available measurements by employing running-coupling prescriptions that mimic the suppression of the spin. Inspired by a discussion in \cite{Kovchegov:2021lvz}, the daughter-dipole prescription is employed in a recent global analysis \cite{Adamiak:2023yhz} comparing the KPS-CTT evolution at large-$N_c\& N_f$ limit to the polarized DIS and SIDIS data at $0.005\leq x\leq x_0=0.1$ and $1.69\;\mathrm{GeV}^2 \leq Q^2 \leq 10.4\;\mathrm{GeV}^2$ from SLAC, EMC, SMC, COMPASS and HERMES experiments. The observables include $A_1$ and $A_{\parallel}$ for polarized DIS and $A_1^{h}$ for polarized SIDIS, with proton, deuteron and helium-3 targets. For SIDIS, we include the production of charged pions, charged kaons and unidentified charged hadrons. Altogether, 226 data points are available. The upper bound of $x_0=0.1$ is due to the limited applicability of the KPS-CTT equation at large $x$. Determined in \cite{Adamiak:2021ppq}, this choice of $x_0$ is an order of magnitude higher than the usual choice of $x_0\sim 0.01$ in similar analyses \cite{Albacete:2009fh,Albacete:2010sy,Beuf:2020dxl} for the unpolarized case. This is attributable to the term that is resum in each evolution, that is, the factor, $\alpha_s\ln^2(1/x)$, resum in the KPS-CTT equation becomes large at a relatively moderate value of $x$ compared to the factor, $\alpha_s\ln(1/x)$, that is resum in the BK equation.

For our global analysis \cite{Adamiak:2023yhz}, the initial conditions at moderate $x=x_0=0.1$ for the polarized dipole amplitudes are given by linear combinations of rapidity and transverse logarithms, together with the constant terms. More explicitly, 
\begin{align}\label{IC}
&(\mathrm{Dipole})\Big|_{x=x_0} = a\,\ln(\mathrm{rapidity}) + b\,\ln(\mathrm{transverse\;dipole\;size})+c\,.
\end{align}
This form is a generalization that includes all the terms that appear in the polarized dipole amplitudes at Born level. With different types of sub-eikonal exchanges and different flavors of quarks, the setup includes 5 different polarized dipole amplitudes, each of which has a distinct set of 3 parameters. In total, the estimates of the 15 free parameters are determined by the global analysis \cite{Adamiak:2023yhz} that employs the data set discussed above.

The analysis is performed within the JAM Monte Carlo Bayesian framework \cite{Sato:2016tuz}. The resulting fit has $\chi^2$ of 1.03 per degree of freedom, implying that the KPS-CTT equation is capable of accurately describing the polarized DIS and SIDIS world data \cite{Adamiak:2023yhz}. Furthermore, the continuation of the theoretical prediction into smaller-$x$ region contains much lower uncertainty than the DGLAP-based approach when coupled with EIC pseudo-data. 

Another main result of the global analysis is the prediction of parton spin contribution from the small-$x$ region. Given the estimates and uncertainties of the free parameters in the initial conditions, we find that
\begin{align}\label{spin_est}
&(S_q+S_g)\Big|_{\mathrm{small}\;x} \simeq \int\limits_{10^{-5}}^{0.1}dx \left[\frac{1}{2}\Delta\Sigma(x) + \Delta G(x)\right] = -0.64\pm 0.60
\end{align}
at $Q^2 = 10$ GeV$^2$. We see that there is likely a significant spin contribution coming from the small-$x$ region. Although the total spin estimate is large and negative, the uncertainty is also large. The latter is a result of error propagation from the parameter estimates. This implies that the model employed for the initial condition contains a large number of free parameters relative to the amount of data currently available. Potential remedies include (i) a more constraining model for the initial conditions \cite{ValQkPol} and (ii) adding more observables to the global analysis. At the time of writing, there are ongoing studies in both directions.

Last but not least, the issue of large uncertainty is shown to significantly subside once the relevant EIC measurement becomes available \cite{Adamiak:2023yhz}. This demonstrates the way the KPS-CTT evolution equation is expected to work hand-in-hand with the upcoming EIC to dramatically extend our understanding of parton helicity in the region of small Bjorken $x$, which is crucial to the complete resolution of the proton spin puzzle.

\section{Conclusion and Outlook}\label{sect:conc}

In this article, we outline the setup and derivation of the KPS-CTT evolution equation \cite{Cougoulic:2022gbk}, which described parton helicity at small Bjorken $x$, and recap most important observations from the recent global analysis \cite{Adamiak:2023yhz} comparing KPS-CTT evolution with the world data of polarized DIS and SIDIS. At this point, the consistency has been established between the KPS-CTT evolution with running coupling and experimental measurements, although the end result, which is the prediction of total parton helicity at small $x$, still contains a large uncertainty. 

To address this issue, a calculation of an alternative model for the polarized dipole amplitude at moderate $x$, which is the initial condition for the KPS-CTT equation, is in progress \cite{ValQkPol}. Inspired by the approach in \cite{Dumitru:2018vpr,Dumitru:2020fdh,Dumitru:2020gla} for unpolarized dipole amplitude, the model estimates the proton state at moderate-$x$ by a Fock state of three valence quarks. Subsequently, the polarized dipole amplitudes can be constructed by taking the expectation value of the sub-eikonal operators on the valence-quark state, including the perturbative corrections due to emission of one gluon. This approach yields an initial condition with much fewer free parameters, as the excess degrees of freedom will be replaced by a physical description of the proton state. 

Another effort to address the large uncertainty in \cite{Adamiak:2023yhz} is to include into a more comprehensive global analysis particle production measurements from $pp$ collision, which directly probe the gluon hPDF. To this end, the theoretical calculation is available in the pure-glue regime, in which an extension to the global analysis is currently ongoing. Furthermore, the extension of the theoretical calculation to include quarks is also in progress.

As mentioned in the article, the most complete picture of helicity at small $x$, together with its interplay with the unpolarized dipole amplitude and gluon saturation, requires the SLA corrections to the currently available double-logarithmic KPS-CTT equation. Such corrections for some of the sub-eikonal operators have been derived in \cite{Kovchegov:2021lvz}. However, the remaining SLA contributions are still in progress \cite{SLAops}, together with the study of compatibility between the small-$x$ helicity evolution and the polarized DGLAP evolution in the overlapping regime.

Finally, it is worth noting that the framework outlined in this article can be applied to similar problems. Since 2019, the orbital angular momentum of small-$x$ partons inside the proton has been studied under a similar framework outlined here \cite{Kovchegov:2019rrz,Kovchegov:2023yzd}. Furthermore, the sub-eikonal expansion to the Wilson lines in the shockwave picture has been generalized to other spin bases, allowing for the study of other TMDs at small $x$ \cite{Kovchegov:2021iyc,Kovchegov:2022kyy,Santiago:2023rfl}.

\section{Acknowledgments}

The author would like to thank D. Adamiak, N. Baldonado, F. Cougoulic, Y. V. Kovchegov, W. Melnitchouk, D. Pitonyak, N. Sato, M. D. Sievert and A. Tarasov, who are the collaborators for all the work this article is based on. Furthermore, the author would like to thank the organizers of the 52nd International Symposium on Multiparticle Dynamics (ISMD 2023) for the invitation to present and discuss the work of our research program. 

The author has been supported by the Academy of Finland, the Centre of Excellence in Quark Matter and projects 338263 and 346567, together with the European Union’s Horizon 2020 research and innovation programme by the European Research Council (ERC, grant agreement No. ERC-2018-ADG-835105, YoctoLHC), and the STRONG-2020 project (grant agreement No. 824093). The content of this article does not reflect the official opinion of the European Union and responsibility for the information and views expressed therein lies entirely with the author.

\bibliographystyle{JHEP-2modlong}
\providecommand{\href}[2]{#2}\begingroup\raggedright\endgroup

\end{document}